\pdfoutput=1  

\documentclass{article}
\usepackage{graphicx}
\usepackage{wrapfig}
\usepackage{subcaption}
\usepackage[T1]{fontenc}
\usepackage{helvet}
\usepackage{mathbbol}

\usepackage{booktabs}
\usepackage{hyperref}
\usepackage{float}
\usepackage{boldline} 
\usepackage{hhline}
\usepackage{longtable}
\usepackage{iclr2026_conference, times}
\iclrfinalcopy

\usepackage[maxbibnames=1, giveninits=true, sorting=none, style=nature, maxcitenames=1, uniquelist=false, doi=false,
  mincitenames=1,
    maxbibnames=5,
  minbibnames=1]{biblatex}
\DeclareSourcemap{
  \maps[datatype=bibtex]{
    \map{
      \pertype{article}
      \step[fieldset=eprint, null]
    }
}
}

\DeclareBibliographyDriver{misc}{%
  \usebibmacro{bibindex}%
  \usebibmacro{begentry}%
  \usebibmacro{author/editor+others/translator+others}%
  \setunit{\labelnamepunct}\newblock
  \usebibmacro{title}%
  \newunit
  \printlist{language}%
  \newunit\newblock
  \usebibmacro{byauthor}%
  \newunit\newblock
  \usebibmacro{byeditor+others}%
  \newunit\newblock
  \printfield{howpublished}%
  \newunit\newblock
  \printfield{version}%
  \newunit\newblock
  \printlist{organization}%
  \newunit
  \printlist{location}%
  \newunit\newblock
  \usebibmacro{doi+eprint+url}%
  \setunit{\addspace}%
  \printtext[parens]{\usebibmacro{date}}%
  \newunit\newblock
  \printfield{note}%
  \newunit\newblock
  \usebibmacro{addendum+pubstate}%
  \setunit{\bibpagerefpunct}\newblock
  \usebibmacro{pageref}%
  \usebibmacro{finentry}}

\AtEveryBibitem{
  \clearfield{doi}
  \clearfield{url}
  \clearfield{isbn}
  \clearfield{issn}
  \clearfield{pagetotal}
  \clearfield{issue}
  \clearname{editor}
  \clearfield{organization}
  \clearlist{location}
}

\usepackage{xcolor}
\usepackage{enumitem} 
\usepackage{notes2bib}

\usepackage[most]{tcolorbox}

\title{Can AI Follow in Einstein's Footsteps?}

\author{%
Michael Shalyt\textsuperscript{1} \quad
Nathan Regev\textsuperscript{1} \quad
Marin Solja\v{c}i\'{c}\textsuperscript{2} \quad
Ido Kaminer\textsuperscript{1}\thanks{Corresponding author: \texttt{kaminer@technion.ac.il}.}\\
\textsuperscript{1}Technion - Israel Institute of Technology, Haifa 3200003, Israel.\\
\textsuperscript{2}Massachusetts Institute of Technology, Cambridge MA 02139, USA.\\
}

\begin{document}

\maketitle
\vspace{-10pt}
\begin{abstract}

AI is accelerating physics discovery, but perhaps away from Einstein-level theory building.
To understand this gap, we must recognize a striking trend: while being very successful,
the most visible AI contributions to physics discovery
appear to mirror the historical development of physics, but in reverse.
Human discovery in physics progressed, in broad strokes, from ancient pattern prediction, through phenomenological laws such as Kepler's, to principle-based universal theories such as relativity and the Standard Model. 
On the AI side, prominent contributions to physics discovery point in the opposite direction:
early milestones emphasized
explicit equation-discovery methods, such as symbolic regression, 
whereas more recent frontier contributions are powerful predictors such as AlphaFold and GraphCast, which can be remarkably accurate yet do not provide clear theoretical understanding.
If this trend continues, 
AI would become extraordinarily good at prediction but may struggle to ever propose its first serious contender to quantum gravity or other paradigm-level theories.
We review the current landscape of AI for physics discovery and highlight a critical missing skill: 
the ability to pose the right questions or invent the right principles to guide the development of new theories and the tests to falsify them.
This mode of discovery has driven many of the deepest advances since the 17th century, 
where symmetry, simplicity, and new mathematical frameworks guided theory construction before experimental tests.
Equipping AI systems with such skills could move them from predicting within known frameworks to proposing the next paradigm-level discovery in physics.

\end{abstract}

\section{Automation of Scientific Skills}
\label{section-introduction}

In Douglas Adams’ The Hitchhiker's Guide to the Galaxy, the supercomputer Deep Thought famously calculates the ``Answer to the Ultimate Question of Life, the Universe, and Everything'' to be 42. 
Adams captured the idea that real progress in physics often requires figuring out the right question rather than just finding an accurate answer.
Current developments in AI for science, and especially in AI for physics, risk heading toward the Deep Thought conundrum in real life.
The Perspective below reviews the history of advances in AI for physics to highlight a growing emphasis on black-box approaches -- often at the expense of simpler solutions and deeper physical understanding -- before proposing paths forward.

Recent years have shown major strides in the ability of AI systems to accelerate scientific discovery across most fields of science \supercite{Wang2023AIscience, kramer2025automated, udrescu2020aiFeynman}. New algorithms increasingly demonstrate skills previously associated with expert scientists \supercite{kitano2016artificial, Carleo2019MLphysics}. Early milestones include automated proof construction, exemplified by the computer-assisted proof of the four-color theorem \supercite{appel1977solution} and interactive theorem provers such as Lean \supercite{moura2015lean}. Subsequent work simulated creative mathematical reasoning through automated conjecture generation in graph theory (Graffiti \supercite{graffiti}), number theory (the Ramanujan Machine \supercite{Raayoni2021, elimelech2023PNAS}), knot theory \supercite{davies2021advancing}, and matrix multiplication \supercite{fawzi2022discovering}.

In physics, algorithmic tools long ago began automating core components of scientific reasoning. 
\textbf{Pattern-recognition skills}, such as anomaly detection in cosmology and high-energy physics experiments, are now routinely delegated to machine learning systems \supercite{BELIS2024AnomalyParticlePhys}. 
AI systems also automate the construction of predictive \textbf{symbolic formulas from data} with symbolic regression systems such as BACON \supercite{langley1977bacon, langley1987BaconBook}, Eureqa \supercite{Schmidt2009Eureqa}, and SINDy \supercite{brunton2016sindy}. 
Recent advances in this area include the recovery of physical equations from the Feynman Lectures on Physics by the AI Feynman Project \supercite{udrescu2020aiFeynman}, derivation of complex ecological and climate equations by knowledge-guided deep symbolic regression \supercite{li2024KG}, and discovery of an analytical formula candidate for predicting the concentration of dark matter from the mass distribution of nearby cosmic structures \supercite{cranmer2020SymModels}.
Although these and other early AI achievements mimicked human-like pattern recognition, more recent prominent contributions of AI to physics follow a different trajectory \supercite{langley2022computational}.

Neural networks expanded the skill of generalization and no longer search for explicit symbolic formulas: The most prominent success of AI for science to date, protein folding by AlphaFold \supercite{senior2020AlphaFold1, Jumper2021alphafold, abramson2024AlphaFold3}, 
demonstrates that neural networks can achieve \textbf{predictive performance without explicit symbolic formulas}, yet providing accuracy surpassing decades of human effort.
Conceptually similar efforts have increasing impact in additional fields, such as predictive models trained on weather data \supercite{lam2023weather}.

In contrast to the approaches above, when the governing formulas are known, they can help the training process by constraining the learning objective or by shaping the inductive bias of the neural network, as in physics-informed neural networks (PINNs) \supercite{raissi2019PINN, raissi2020hidden}. 
In other settings, established physical formulas can be replaced by learned surrogate models \supercite{Nanophotonic2018Marin, Pestourie2020ActiveLO, EmbedEmulate2022, ELORZACASAS2025109105} that emulate complex dynamics while reducing computational costs.

In all of the above, algorithms have successfully simulated specialized skills.
More recently, large language models (LLMs) and their reasoning variants gradually automate general research skills \supercite{wei2022chainLRM, boiko2023emergentLLMScientific,Bran2024LLMscience,  xu2025largereasoningmodelssurvey}. 
Examples include Olympiad-level mathematical reasoning \supercite{2022Minerva, 2022MiniF2F, 2024AlphaGeometry}, unifying mathematical knowledge representations \supercite{raz2025PiUnifier}, 
autonomous proofs of open questions \supercite{PoluSutskever2020generativelanguagemodelingautomated, 2024DiscoveryGenerative, sothanaphan2026resolutionerdhosproblem728, chen2026felsconjecturesyzygiesnumerical, ma2026erdhosproblemrandomsubset}, and analytical representations of scattering amplitudes in high-energy physics \supercite{guevara2026singleminusgluontreeamplitudes, guevara2026singleminusgravitontreeamplitudes}. 

Although much of the discussion around AI for science treats disciplines uniformly, physics has a unique relationship with AI. Physical platforms such as integrated photonics inspire new hardware architectures for accelerating computation \supercite{shen2017nanophotonicsforDL}. Concepts from statistical mechanics inform research on explainability \supercite{bahri2020statistical}, while other concepts inspire changing model architecture \supercite{toscano2025pinns, 2025KAN}.
The fact that many physical systems are governed by compact, universal equations makes them particularly amenable to approaches such as PINNs, which embed these equations into the learning objective \supercite{raissi2024pinnreview}. Agentic AI robots are being developed to automate experimental skills from building setups to performing experiments \supercite{2020RoboticChemist, boiko2023autonomousChemResearch}. These interfaces are reviewed elsewhere \supercite{karagiorgi2022machine, thiyagalingam2022scientific, krenn2022scientific, gentine2024climate, jiao2024ai, makke2024perspective, wetzel2025interpretable, ray2025generative}.

Essential questions remain:
What skills are required for AI to eventually create its first paradigm-shifting discovery in physics? 
How far are we from an AI-generated competitor to the Standard Model of particle physics or to $\Lambda$CDM of cosmology? 

\textbf{The bottom line of the Perspective} is that
achieving such breakthroughs requires more than better prediction skills: 
It requires AI systems capable of identifying the right questions,
sustaining high-risk hypotheses long enough to test them,
and constructing new theoretical frameworks guided by abstract principles, 
possibly through the development of new kinds of mathematics.

\vspace{-8pt}
\section{The Reverse Epistemic Trajectory}
\label{section-evolution}
\vspace{-8pt}

\begin{figure}[h!]
    \centering
    \includegraphics[width=1\linewidth]{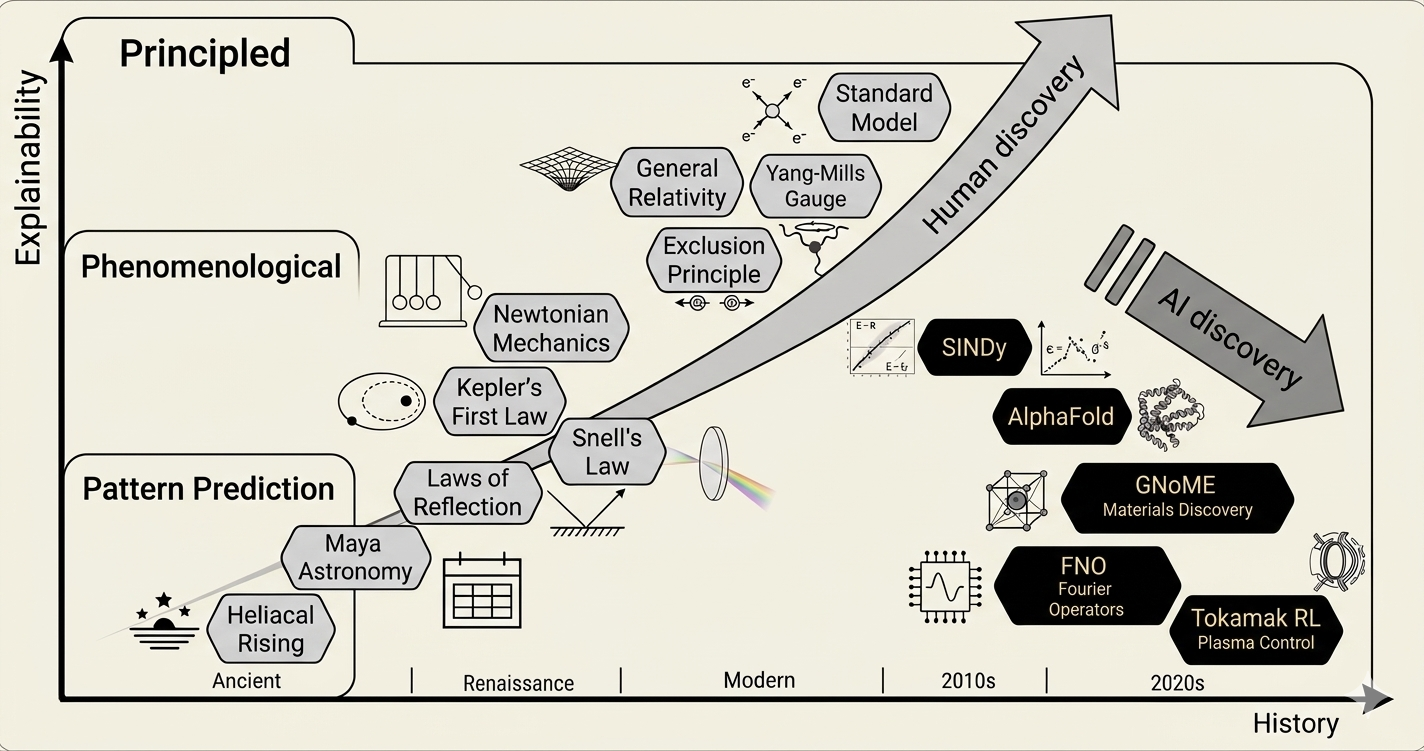}
    \caption{\textbf{The trajectories of human vs AI physics discovery -- reversal in explainability.}
    The arrows summarize example contributions representing an apparent trend.
    Many milestone discoveries throughout history followed a trend: from pattern prediction to phenomenological laws and finally to principle-based theories, the level of ``understanding'' is increasing. 
    In contrast, 
    the center-of-gravity of AI contributions to physics has followed this trend in the reverse direction: from discovery of explicit symbolic formulas to neural network predictive systems with decreasing explainability. 
    }
    \vspace{-8pt}
    \label{fig:human-vs-AI-history}
\end{figure}

There is no single story describing the discovery process in the history of physics, but until recently, it followed a trajectory towards deeper levels of understanding. AI is now following a similar trajectory, but apparently in reverse epistemic order (Fig.~\ref{fig:human-vs-AI-history}).

The scientific revolution established a new standard for understanding nature: derivation of universal mathematical laws from empirical observation, as Newton famously framed in the \textit{Principia} \supercite{newton1687philosophiae}. His achievement was to identify mathematical laws that govern phenomena across scales, from falling apples to orbiting planets. 
Since then, the formulation of universal laws from empirical observation has been regarded as the hallmark of physical ``understanding''.
From the perspective of AI, symbolic regression aims to automate precisely this process: finding mathematical laws from empirical data. 
Yet more recent developments in AI for science, such as GNoME \supercite{merchant2023gnome}, provide accurate predictions without explicit mathematical laws, echoing physics discoveries before the scientific revolution.
It is in this sense that AI discovery in physics is following the history of human discovery in reverse order.

This reversal is not a universal law of AI for science, nor a chronology of AI architectures. 
Rather, it reflects human incentives and practical choices that have recently shifted the field’s center of gravity: 
many of the most visible and influential AI-for-science successes are now produced by systems optimized for predictive power rather than for discovery of explicit, human-interpretable theories. 
These successes coexist with -- but increasingly overshadow -- 
efforts to make AI discover organizing principles and mathematical structures, and thereby become a participant in theory building.

\vspace{18pt}

\vspace{-2pt}

At a coarse level, the history of physics may be viewed as progressing through three broad stages:
\textbf{Pattern Prediction -} Ancient systems, such as the Metonic cycle or Mayan astronomy, predicted celestial events with high accuracy via cycles and correlations, without explicit mathematical laws.
\textbf{Phenomenology -} Kepler's three laws and Planck's early black-body radiation formula provided mathematical equations that fitted the data but lacked the underlying principles to explain it.
\textbf{Principle-Based Theory -} Newton, Maxwell, Heisenberg, Schrödinger, Einstein, and others established fundamental principles and a new mathematical language, from which specific laws could be deduced.

The development of successful AI contributions to physics has mirrored the two earlier stages of human physics, but in reverse order.
Early symbolic-regression models such as BACON, Eureqa, and SINDy extract equations from data -- akin to discovery of phenomenological laws. 
More recent AI systems based on neural networks \supercite{lam2023weather} increasingly abandon explicit equations and function as high-accuracy predictive oracles -- closer in spirit to pre-scientific pattern prediction.

This historical sketch is necessarily approximate, not a strict historical timeline. One caveat is worth highlighting.
In the eyes of some ancient scientists, their theory was also principle-based, 
even if their principles were often metaphysical rather than based on mathematical laws.
Ptolemaic astronomy, for instance, aimed to represent celestial motion guided by abstract principles about cosmic harmony and the perfection of idealized circular motion. 
Kepler rejected the idea that his laws are mere calculational devices fitting Tycho Brahe's data, instead explicitly framing his seminal book Astronomia nova as ``celestial physics'', seeking principles he regarded as genuinely explanatory.
They did not see their work as phenomenological.
In contrast, certain human discoveries that are seen today as symbols of principle-based advances were discovered without those principles, or even using the wrong ones. 
Maxwell’s route to field theory initially relied on mechanical scaffoldings (``molecular vortices'') that were later discarded while the equations survived \supercite{maxwell1861physicallines1}. 

This caveat does not overturn the broader trend. 
What matters in modern view is not whether historical actors believed they had principles, but whether the resulting discovery compressed more phenomena into a simpler and more powerful mathematical structure. 
That is the kind of progress exemplified by relativity, quantum theory, and gauge theory. 
In contrast, much of current AI-driven discovery improves prediction while retreating from the desire for mathematical simplicity. 
What recent history reveals is a shift in ideals and ``optimization target'' regarding AI physics discovery --
from prediction with understanding to prediction without it.

\section{AI Prediction Without Understanding}
\label{section-prediction-without-understanding}

The shift of many high-impact AI applications toward black-box surrogates reactivates
a pre-scientific-revolution mode of discovery, albeit one operating with superhuman efficacy. 
A growing number of frontier AI for physics systems prioritize utility over the use of explicit mathematical laws, and indeed provide a lower level of (human) understanding \supercite{messeri2024artificial, langley2024discovery}.
This prioritization provides an immediate benefit in predictive payoff.

Certain problems in physics still benefit from a partial reliance on explicit formulas and laws, as common
in PINNs \supercite{raissi2019PINN}. These AI models embed known physical laws, typically partial differential equations (PDEs), directly into the network's loss function to act as structured regularizers 
\supercite{cai2021physics, karniadakis2021physics}.
Similarly, design priors can be used to enforce exact physical constraints \supercite{loh2022surrogate}, 
such as embedding Lagrangian mechanics to guarantee energy conservation \supercite{cranmer2020lagrangianneuralnetworks}.
While such neural networks are highly effective, their scope is still limited to particular domains, as they require an externally-supplied theoretical foundation -- which is not discovered by the networks.

In contrast, increasingly popular approaches now use purely black-box surrogates such as Google DeepMind’s GraphCast \supercite{lam2023weather} and NVIDIA’s Project NeRD \supercite{xu2025NeRD}.
Rather than enforcing known physical laws (e.g., Navier-Stokes equations), these surrogate systems treat complex phenomena like fluid simulations as high-dimensional statistical prediction problems \supercite{sanchez-gonzalez2020learning}. By training on massive datasets, these systems often surpass the accuracy of traditional numerical solvers, yet they do so without an explicit representation of the fluid dynamics or mechanical principles involved.

Historically, this predictive surrogate approach resembles ancient astronomical traditions such as the Antikythera mechanism \supercite{seiradakis2018current} or Mayan calendrical systems. These were masterpieces of forecasting, capable of predicting eclipses and planetary positions with high precision, yet they possessed no understanding of gravity or orbital mechanics. 

\begin{figure}[h!]
    \centering
    \begin{minipage}{0.55\linewidth}
        \centering
        \includegraphics[width=\linewidth]{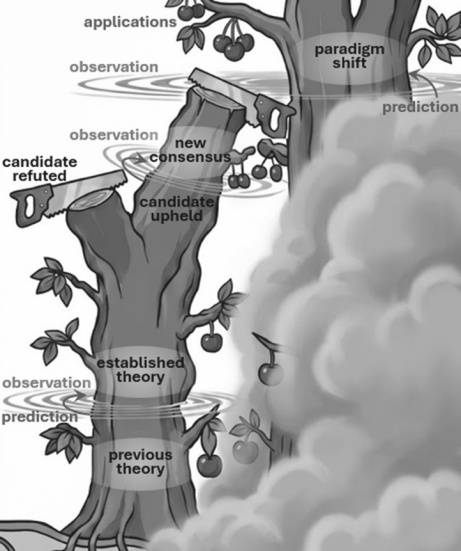}
    \end{minipage}
    \hfill
    \begin{minipage}{0.44\linewidth}
        \caption{\textbf{Evolution of scientific theories.}
        A schematic representation of the iterative, upward growth of scientific understanding. Every extension of an established theory provides predictions that are tested against observations, either refuting this theory extension (tree stump) or upholding it (continued growth of the trunk). Every extension of the theory provides further applications (fruits) and increased understanding. Paradigm Shifts, in the sense of Kuhn, could be considered as the sudden revelation of a separate tree providing the previous applications and understanding, as well as new ones that cannot be otherwise accessed. We ask what developments would enhance the ability of AI systems to mimic the entire iterative process, and potentially even assist the discovery of unexpected paradigm shifts.}
        \label{fig:theory-evolution}
    \end{minipage}
    \vspace{-10pt}
\end{figure}

From the broader perspective of the AI community, the fact that a model does not mirror the underlying physics is no longer viewed as a problem.
This shift in priorities can be attributed to the Bitter Lesson \supercite{sutton2019bitter, 2024learning_Bitter_Lesson} of AI development: methods that leverage massive computation and data often outperform those relying on human-designed theoretical structures. 
In this view, the value of a model lies in whether it predicts accurately and efficiently and not in whether it recovers the true mechanism.
In fact, a model can be extremely useful even if its internal ``understanding'' is physically wrong. 
An analogous situation appears in the history of optics. 
The optical theories of Euclid, and later Ptolemy and Ibn al-Haytham, produced quantitatively successful descriptions of refraction and the passage of light through materials, despite operating on fundamentally incorrect premises regarding the nature of vision.
Neural network black-box models recreate this situation almost inevitably. They are not optimized to recover the true underlying mechanism, but to compute accurate outputs with an efficient universal architecture based on matrix multiplications and nonlinear activation functions. 
This architecture can often accurately predict the data  at lower computational cost or greater speed than explicit mathematical laws, while being oblivious to the true underlying equations.

But we should not let this apparent advantage lead us to abandon the search for simpler, explicit theories. 
The Bitter Lesson explains why black-box predictors can dominate in data-rich regimes. Yet paradigm-level physics often begins in the opposite regime, where data is sparse or conceptually misinterpreted. In such cases, historical progress has often depended not on prediction alone, but on hypothesized mathematical structures that 
expose new variables, constraints, and principles.

This is precisely the limitation identified by
Bunge in ``A General Black Box Theory'' \supercite{bunge1963blackbox}: a theory that deals solely with the relationship between inputs (stimuli) and outputs (responses) lacks ``conceptual handles'' that can be interrogated or proven wrong. A paradigm shift, in the Kuhnian sense \supercite{kuhn1962structure}, requires a crisis within the underlying theory; however, a black-box predictor with no explicit structure or epistemological interpretation cannot undergo such a crisis -- it can only be re-optimized.
If that AI model then undergoes a re-optimization granting it unprecedented predictive capabilities, on the level of a paradigm shift in their application -- would it count as a true paradigm shift?

\begin{tcolorbox}[
colback=blue!8,
colframe=blue!60!black,
title=\textbf{Caveat (``Ipcha Mistabra''): Do we Really Need to Understand AI Discoveries?
},
fonttitle=\bfseries,
boxrule=1.2pt,
boxsep=3pt,
left=4pt,
right=4pt,
top=4pt,
bottom=4pt]
There are reasons to believe AI would also be better if it was always trying to come up with simplified, explainable representations of the world around it.  
But this is not necessarily so, 
and a provocative possibility is that some AI discoveries will rely on new types of mathematics that the AI will not be able to explain to humans but be able to communicate at a higher level (e.g., between AI agents) -- so attempts at human understanding might just slow down progress.
\vspace{2pt}

Related phenomena have already appeared in studies of emergent communication, where cooperating agents can drift away from natural language toward task-specific communication protocols \supercite{chaabouni2022emergent, lee2019countering}. 
More recently, LLMs have begun to explore reasoning in continuous latent spaces rather than through explicit natural-language chains of thought \supercite{hao2025traininglargelanguagemodels}. 
If such systems were to generate powerful but non-human-interpretable representations, a central challenge would become how to recognize when a significant discovery had occurred at all (potentially by quantifying compression or complexity of the predictive structure).

\vspace{2pt}
Could such an advance be considered a paradigm shift?
Whatever status one assigns to non-human-interpretable advances, they do not remove the need for explicit, falsifiable theory in physics. 
The central agenda of this Perspective is therefore to argue for AI systems that do more than predict: 
they should help formulate principles, identify the right questions, and direct the experiments that can test them.

\vspace{2pt}

\end{tcolorbox}

\textbf{Explainability in physics for AI alignment.} AI models that predict without explaining how they do so are harder to supervise, falsify, verify, map their limitations, and eventually trust.
Such trust is becoming vital as practical applications increasingly rely on black-box predictions and results. 
Even if alignment and accuracy are ensured by other means, there remains a need for human-level explainability to provide feedback and direct the AI research towards types of discoveries that so far remain inaccessible by AI systems.
Shifting AI development from prediction without understanding to enhanced explainability is a prerequisite for future human-AI collaboration.

\section{Categories of Physics Discoveries: Human vs AI} 
\label{section-categories-physics-discoveries}

The deepest scientific breakthroughs often require more than new equations: they introduce new organizing principles, and sometimes entirely new mathematical language.
In fundamental physics, especially over the past century, many paradigm shifts have come precisely from this move toward new mathematical frameworks.
This thinking motivates the classification of discoveries in physics into three rough categories.
These categories allow us to verbalize the epistemological leaps required for discovery, 
and do not serve as strict mutually exclusive classifications.

\begin{tcolorbox}[colback=blue!8,colframe=blue!60!black,title=\textbf{Sorting Physics Discoveries by Levels of Mathematical Abstraction},fonttitle=\bfseries, boxrule=1.2pt, boxsep=3pt, left=3pt,right=3pt,top=4pt,bottom=4pt]

\begin{center}
\footnotesize
\setlength{\tabcolsep}{3pt}
\begin{tabular}{p{0.11\linewidth} p{0.27\linewidth} p{0.27\linewidth} p{0.27\linewidth}}
\toprule
\textbf{Category} & \textbf{Category A} & \textbf{Category B} & \textbf{Category C} \\
\midrule[0.7pt]

\textbf{Type} &
new solution or capability &
new equation or law &
new math framework \\[4pt]

\midrule[0.1pt]

\textbf{Description} &
prediction, device, application &
theory, ansatz, formula &
mathematical language, symmetry principle\\

\midrule[0.1pt]

\textbf{Human \newline Discovery \newline Examples} &
ballistics and orbital prediction; photonic devices and antennas; transistors and lasers; collider phenomenology&
Kepler's laws; Lorentz force and the Biot-Savart Law; semiconductor band theory and stimulated emission; Standard Model and Higgs mechanism&
Newtonian mechanics and calculus; field theory and vector calculus; Hilbert space and operator formalism; renormalization and gauge theories in quantum field theory\\[4pt]

\midrule[0.1pt]

\textbf{AI \newline Discovery \newline Examples} &
Solid and fluid material simulations using graph networks; protein folding and weather prediction using transformers; data-efficient approximators for problems in quantum mechanics and hydrodynamics using PINN&
analytic formula for dark matter concentration using symbolic regression; analytical representations of scattering amplitudes by LLMs; ``inverting'' the process of effective field theory by computational means&
\begin{center}
\textbf{?}
\end{center}
\\

\bottomrule
\end{tabular}
\end{center}

To contribute to paradigm shifts, AI should move beyond making predictions and even beyond discovering new solutions and equations, toward inventing new mathematical frameworks.
\end{tcolorbox}

These categories also apply outside of physics. A classic example is the discovery of formulas for fundamental constants such as $\pi$ and $e$ by mathematicians over the centuries (Category A), many of them found to be special cases of the theory of continued fractions (Category B), recently found to be captured and unified by the mathematical framework of conservative matrix fields (Category C) \supercite{elimelech2023PNAS, raz2025PiUnifier}.

Nevertheless, we should not expect every discovery to fall neatly within one of the categories A-B-C. General relativity is clearly Category C, whereas the Schwarzschild metric is technically a particular solution and thus closer to Category A. Yet some solutions have consequences so profound -- revealing black holes, horizons, singular behavior -- that they reshape how we understand the framework from which they arise, and even point the way toward future Category C theory developments, such as modern efforts in quantum gravity.

This blurring of categories points to the core issue: the most profound gap in current AI for physics discovery is not the absence of Category C outputs, but its inability to replicate the mode of discovery that produced many of the great advances of 20th-century physics.
Current AIs (including LLMs and symbolic regression) excel at \textit{induction}: evaluating how well a hypothesis matches the data. They also possess growing capabilities in \textit{deduction}: computing the consequences of a given physical theory or framework (via simulation or logic).
However, many great leaps of modern physics (e.g., general relativity, the Dirac equation, the Higgs mechanism) were neither inductive nor deductive. Einstein did not build general relativity using data about Mercury's orbit. He started with abstract principles of symmetry (equivalence, covariance), \textit{abducted} a candidate theory satisfying those principles, and only then sought empirical verification. 
This is the Popperian ideal: to construct bold theories, deduce empirical predictions, and subject those predictions to experimental tests aimed at falsifying them.
This discovery pattern seems not to have been captured yet by current AI systems for physics.

\begin{tcolorbox}[colback=blue!8,colframe=blue!60!black,
title=\textbf{Historical Examples of Principle-Guided Human Discoveries},
fonttitle=\bfseries, boxrule=1.2pt, boxsep=3pt, left=3pt,right=3pt,top=4pt,bottom=4pt]
\small

\begin{tabular}{p{0.47\linewidth} p{0.47\linewidth}}

\textbf{Laughlin's theory for the fractional quantum Hall effect (Category A)}
found an explicit many-body wavefunction structure fixed by symmetry and simplicity in the lowest Landau level \supercite{Laughlin1983}; deduced incompressibility and fractionally charged excitations, capturing the essence of a new quantum fluid.

&
\textbf{Bethe ans\"atze (Category A)}
solved the one-dimensional spin chain by guessing a constrained eigenstate form and enforcing consistency conditions on the phases \supercite{Bethe1931}; strong constraints on translation invariance and particle exchange converted the problem into closed-form equations.

\\
\\

\textbf{Ginzburg--Landau theory of superconductivity (Category B)}
introduced an order-parameter field and free-energy functional consistent with symmetries \supercite{ginzburg1950theory}; postulating the right variables and invariances predicted macroscopic superconducting behavior, later shown to work near $T_c$ \supercite{Gorkov1959}.

&
\textbf{Parisi's theory of spin glasses (Category C)}
presented replica-symmetry breaking for disordered systems \supercite{Parisi1979}, replacing a single macroscopic order parameter by a hierarchy of overlaps between pure states; resulting representation of equilibrium state space predicted many universal properties of complex systems.

\end{tabular}

\vspace{8pt}
These physics discoveries all seem to rely on ``well-educated guesses'', which have a shared structure.
Even when based on established frameworks, they all impose an additional structured hypothesis, such as a certain ans\"atze, a symmetry, or gauge -- enforcing simplicity before deducing falsifiable consequences. 
\end{tcolorbox}

\section{Can AI Automate the Rare Insights of Top Physicists?}
\label{section-do-Einstein}

\textbf{Could it be that AI is fundamentally unable to achieve human-level creativity?}
While this had been a long-standing question, a growing body of evidence suggests that there is no fundamental limit on AI creativity \supercite{Si2024LLMCreativity}.
Among many examples, especially visible milestones include DeepMind's highly unconventional ``Move 37'' by AlphaGo \supercite{silver2016Go} that upended centuries of human expert intuition,
and OpenAI's disproof of Paul Erdős's famous planar unit-distance conjecture by an internal model \supercite{alon2026remarksdisproofunitdistance} that combined techniques from two different fields of mathematics.
An even more recent example is Anthropic's counterexample to the Jacobian conjecture by Fable \supercite{Alpoge2026Jacobian} that provided an initially opaque construction, subsequently reverse-engineered and generalized to human-understandable geometric terms \supercite{alexis2026jacobian_explained}. 
These examples represent cases in which AI systems went beyond prior knowledge and made unexpected choices, which created new opportunities for subsequent human understanding.

If the gap is not creativity, 
what missing skills, then, prevent AI from making Category C discoveries? 
Unlike human physicists that have successfully driven progress across all three categories, 
AI-powered breakthroughs to date remain concentrated in Categories A and B, 
with the most visible recent successes increasingly falling on the predictive, low-explainability side of Category A -- a trend that shows no signs of slowing.

AI for physics of Category A solves physics problems or predicts the outcome of physical processes. Such AI systems include surrogate models trained on simulated data, where an expensive simulator is replaced by an efficient neural network \supercite{ziv2025MotiUnsupervisedQMB, 2025NathanScintilators}. When reliable simulations are unavailable, too expensive, or insufficiently accurate, neural networks can instead be trained directly on experimental data. Both simulation-trained and experiment-trained approaches can be enhanced using known physical formulas to improve training-data efficiency, robustness, and generalization.

AI for physics of Category B provides an explicit theory or formula.
Such AI systems include symbolic regression techniques usually building on experimental data. Other approaches rely on symbolic computation to make analytical predictions in search for more general theories consistent with experiments. These approaches are now enhanced by LLMs.

Strikingly, no AI has made a discovery of Category C, involving the successful use of a new principle or the invention of a novel mathematical framework.

\textbf{Why is it so hard?}
The creation of a Category C discovery presents a challenge that current AI models are ill-equipped to meet. 
The primary bottleneck is not just producing a novel ``creative'' idea.
Novelty is cheap in a sufficiently large search space.
The harder task is estimating the scientific payoff of each novel idea:
judging how hard it is to execute, whether it is likely to become useful, and whether it yields falsifiable tests.
This is where a payoff-per-effort ratio matters.
Many top-tier human researchers possess a rare ``gut feeling'' for this selection problem -- a talent for identifying research questions and elegant approximations that are both profound and practically executable. 
Part of this ability comes from experience, and part of it may be a rare form of scientific taste.
Because this ``gut feeling" is such a unique talent and is only rarely written down as a procedure, current AI systems have little direct training data for it.
It is therefore unsurprising that current top AI models struggle to imitate the kind of judgment behind Category C discoveries. 
As long as we cannot train AI systems to imitate this undocumented feeling directly, our leading approach to AI-assisted progress in physics remains to build AI systems that can generate, test, and refine candidate principles until some acquire the marks of a useful theory.

While it remains unclear how Category C breakthroughs could be fully automated, 
the following section sketches several strategies 
for steering AI for physics toward this higher-level mode of discovery.

\section{How to Teach AI Better Modes of Discovery?}
\label{sec:principle-guided-building}

Several complementary strategies could help AI systems move toward Category C physics discoveries. This section is not intended as a comprehensive survey or a closed roadmap; the field is developing rapidly, and new approaches will likely emerge that are difficult to anticipate today. 
Rather, we highlight representative directions that seem especially relevant for principle-guided theory building and for raising the level of understanding and explainability in AI for physics discovery.

\textbf{Propose questions rather than answers.}
As famously shown in the Deep Thought story, it is (much) more important in physics to figure out the right question rather than merely finding the answer. Indeed, the formulation of the question is rarely a simple preliminary step -- it often constitutes the discovery itself. 
Agentic AI systems can be trained to prioritize proposing hypotheses and research questions, then running closed-loop simulations to stress-test and simplify their proposals.
In AI models already used for physics discovery, these exploratory preferences can be reinforced via modular skill definitions (such as markdown-based skill specifications \bibnote{See for example the \texttt{skills.md} repository https://skills.sh}) and eventually integrated into a unified agent harness applying additional cognitive styles proposed below.

\textbf{Training to the tail and developing persistent ``obsessions''.} 
Current generative AI is architected to optimize for mainstream human thought (the statistical mode) rather than outliers (the statistical tail). 
Through next-token prediction and reinforcement learning from human feedback (RLHF), LLMs are trained to smooth away ``weirdness'' in favor of the plausible consensus. Yet, paradigm-shifting breakthroughs are, by definition, radical outliers. 
Human discovery in physics has historically often relied on a capacity for persistent ``obsession'' -- a cognitive style that allows a researcher to hold onto an outlier idea over years. 
Einstein chewed on relativity for a decade; Bednorz and Müller went against prevailing expectations for years before discovering cuprate superconductors.

In comparison, leading AI models are currently trained to produce outputs that look like good thinking \textit{from inside the consensus} of a field. 
Revolutionary work, by definition, looks wrong from inside the consensus until it changes that consensus.
LLMs are optimized to do the exact opposite: notice when the field disagrees and update toward the center. 
To move beyond this, we must build architectures capable of proposing anti-consensus positions and sustaining them over time, while guiding searches to gradually accumulate evidence that can support or refute these positions.

\textbf{Striving for simple interpretable theories.}
There are strong reasons to believe that AI for physics would be more successful if trained to derive simpler theories and explicit equations from the provided data. 
The inherent advantage of such explicit structures is evident in recent works:
Incorporating explicit laws into the AI model as inductive bias enables training with less data \supercite{ZHU2019SurrogateNoLabel, SUN2020SurrogateFluid, ma2025HeuristicRoles} and, more importantly, improves generalization capabilities \supercite{2021MarinRegressionDLIntegration, ma2023topogivity, ma2025BandGap}.

Therefore, alongside prediction-driven AI models that will continue to be developed aggressively, we should invest in AI systems that search for lower-complexity theories. 
This approach is important even for the discovery of \textit{approximate laws}, which remain scientifically powerful even when they are not exact: they can reveal the dominant variables, improve robustness, and support out-of-distribution generalization. 
More generally, whenever a simpler theory exists beneath a complex dataset, finding it is valuable not only for human understanding but also for automated extrapolation and generalization.
The practical target is not only elegance for its own sake, but compression that makes the theory easier to use or test, and its ideas easier to extend or transfer across domains.

\begin{tcolorbox}[
colback=blue!8,
colframe=blue!60!black,
title=\textbf{The Simplicity of Physical Laws: Guiding Star or Trap?},
fonttitle=\bfseries,
boxrule=1.2pt,
boxsep=3pt,
left=4pt,
right=4pt,
top=4pt,
bottom=4pt]
\textit{``The miracle of the appropriateness of the language of mathematics for the formulation of the laws of physics is a wonderful gift which we neither understand nor deserve.''} Wigner 1960 \supercite{wigner1960unreasonable}

\vspace{5pt}
One possible explanation for the ``gift'' Wigner described is the separation of scales. Natural phenomena containing behaviors on different scales of energy, distance, or time, often admit compact effective descriptions in terms of fewer parameters. Historically, this separation enabled discovery of laws one scale at a time, a process that guided many discoveries of fundamental laws of physics over the years.

\vspace{5pt}
But the historical success of physics should not make us assume that an elegant theory lies underneath all domains of physics and other fields of science. For one, the electronic description of matter contains contributions that often cannot be separated in scale, causing inherent complexity in condensed matter physics. Similar challenges are fundamental to many domains of chemistry and biology.

\vspace{5pt}
Where the scales cannot be separated, 
the search for an elegant underlying theory may itself be a hopeless pursuit. 
In such domains, accurate AI predictions without transparent explanation may in fact be the best one can hope for.
The question is therefore not whether explainability is always preferable, but when the lack of explainability signals an unfinished theory and when it reflects genuine irreducible complexity.

\vspace{5pt}
With the rise of AI-assisted discovery, learning to make this distinction becomes essential: 
when should we accept predictive accuracy as the best available outcome, and when should we continue searching for a simpler underlying theory? 
In the case of the success of AlphaFold: is the lack of a closed-form formula for protein folding inherent to this process, or have we just missed the underlying structural laws?
Either way, the predictive success of recent AI models should not tempt us to stop questioning whether a simpler structure remains hidden underneath.
\end{tcolorbox}

\textbf{Enhancing theory building using symbolic computing.} 
To move AI from prediction without understanding to the invention of new principle-based physical frameworks, AI systems require an environment where their abductive ``guesses'' can be rigorously tested. Such an environment should be designed to cultivate in AI the aforementioned ``gut feeling'' of top-tier human researchers. Symbolic computing tools could be used for this purpose, providing \textit{theory-building sandboxes}. 
Computer algebra systems (CAS) could enable leading AI models and agentic LLM systems to write and execute code relevant to theory building in physics.
AI agents connected to symbolic engines can manipulate exact mathematical objects, ensuring that a proposed theory is structurally coherent before it is tested against data.

Despite the potential of this neurosymbolic approach, efforts in this direction have so far been limited (with most examples being in AI for mathematics \supercite{funsearch, alphaevolve}).
Pioneering efforts have used symbolic computing tools to insert symmetries into AI pipelines. 
For instance, recent frameworks have demonstrated how to enforce or discover continuous symmetries via specialized regularizations based on Lie derivatives \supercite{otto2025symmetryML}, 
embed fundamental space-time symmetries directly into network structures \supercite{Gong2023}, 
or integrate theoretical axioms into hypothesis generation to simultaneously satisfy data and physical laws \supercite{CoryWright2024}.
Broader efforts in this direction utilize symmetry constraints in symbolic regression algorithms \supercite{Chen2025Governing, 2025SRBeyondSM}.

Existing methods primarily treat symmetry as a constraint for data fitting or model training. 
There remains a crucial step of ``symmetry abduction'' -- toward principle-guided theory building, requiring development of closed-loop neurosymbolic architectures. 
We see early prospects of this concept in high-energy physics, where transformers have been trained on data generated by advanced symbolic tools to complete missing integer structures in scattering amplitudes \supercite{CaiDixon2024Tranforming, Cai2025RecurrentFeatures}.

Looking forward, AI must move from imposing symmetries and predicting amplitudes to proposing the Lagrangians themselves. A concrete route toward this goal is to convert theory-building into a constrained algorithmic search problem. Early efforts in this direction have been developed in the Standard Model effective field theory (SMEFT). In SMEFT, candidate theories of ``new physics'' are encoded through the most general higher-dimensional operator expansion consistent with gauge symmetries, organized by a strict simplicity criterion (power counting in $1/\Lambda$) \supercite{Grzadkowski2010,BrivioTrott2019}. 
This approach has provided a consistent language for combining empirical constraints with beyond-Standard-Model scenarios \supercite{EllisMurphySanzYou2018}.

The physics community has already built part of the required CAS infrastructure.
For example, effective field theory (EFT) matching is increasingly automated via tools such as \texttt{matchmakereft} \supercite{carmona2022matchmakereft} and \texttt{Matchete} \supercite{fuentes2023matchete}.
Building on this logic, recent computational work has demonstrated the inversion of the usual EFT workflow: starting from a target low-energy (IR) Lagrangian and systematically searching for high-energy (UV) theories whose lower energy limits match it \supercite{LifshitsBUILD}.
By automating part of the search over symmetries, gauge groups, and field properties, this approach illustrates how symbolic infrastructure can support the 
discovery of new candidate theories in domains such as quantum gravity.

To achieve this at scale, AI systems should be trained to use established symbolic tools in these areas of physics. Future ``AI physicists'' could be trained to manipulate tensors and symmetries via \texttt{xAct} \supercite{xAct} and \texttt{Cadabra} \supercite{Peeters2007Cadabra, Peeters2018Cadabra2}, process large symbolic expressions in \texttt{FORM} \supercite{Vermaseren2000FORM, Kuipers2013FORM}, automate quantum many-body derivations in electronic-structure theory with \texttt{SeQuant} \supercite{Gaudel2026SeQuant}, and handle Feynman diagrams using \texttt{FeynCalc} \supercite{ShtabovenkoMertigOrellana2020FeynCalc} and \texttt{FeynRules} \supercite{Alloul2014FeynRules2}.
By integrating such CAS code into agentic workflows, we can create the desired \textit{theory-building sandbox} for AI systems. 
In it, the AI can generate novel mathematical frameworks, filter out redundant or mathematically inconsistent ideas, and optimize for the rare, high ``payoff-per-effort'' insights that characterize the greatest human discoveries.

\textbf{Artificial intuition via physics-aware world models.}
Human physicists are constantly driven to compress complex observations into simplified, consistent mental representations of reality. 
This internal ``world model'' is a source of physical intuition, allowing scientists to filter out contradictory hypotheses and identify elegant approximations. 
This physical intuition is instrumental for the construction of model-based 
\textit{Gedankenexperiments} that famously guided the development of new domains of physics.
To develop a similar intuition in AI, there are strong reasons to believe we must architect systems that are explicitly forced to come up with simplified representations of their environment. 
Developing an artificial world model \supercite{lecun2022path} that learns the underlying dynamics of how the world evolves could be the key to granting AI genuine physical intuition.

Currently, efforts toward learning a world model are highly fragmented. Separate systems achieve superhuman accuracy in protein folding, global weather prediction, etc. 
It may be that each task-specific system already holds an underlying intuition of thermodynamics or fluid dynamics. 
This intuition could be shared among AI systems by direct communication between AI systems, or by one AI using others as tools.

A promising direction for overcoming this fragmentation lies in training scientific foundation models on massive, heterogeneous corpora of direct physical measurements -- encompassing microscopic, spectroscopic, astronomical imaging, and sensor data. Early indicators of this broader approach are emerging in AI models designed to unify multiple scientific data types within a shared learning framework \supercite{bodnar2025aurora}.

The hypothesis is that scaling up multi-domain physical data will force the AI system to find transferable structures or principles common across seemingly different domains of physics. 
If an AI is forced to predict both the Newtonian trajectories of planetary motion and the effect of gravitational lensing using the same latent space, 
it may be pushed to find a unifying framework, e.g., Einstein's relativity in this case.
The hope is for such unification to lead to a compressed representation that can be translated back into explicit equations and human-understandable theory.

\begin{tcolorbox}[
colback=blue!8,
colframe=blue!60!black,
title=\textbf{Is the Bottleneck AI, or Physics Itself?},
fonttitle=\bfseries,
boxrule=1.2pt,
boxsep=3pt,
left=4pt,
right=4pt,
top=4pt,
bottom=4pt]
Some may argue that AI has failed to find the next general relativity because humans may not find one either: perhaps the era of simple, experimentally testable theoretical revolutions in physics is over. The past few decades could be read as evidence for this view.
\vspace{4pt}

We believe it is too early to announce the end of theoretical physics. 
This concern should instead inform how we test AI systems for physics discovery. 
We can give AI systems artificial worlds with hidden laws and test whether they can invent simple theories inside those worlds. 
Alternatively, we can design historical reconstruction tasks such as Demis Hassabis’ ``1911 cutoff'' test \supercite{officechai2026hassabis1911}, recently attempted in ``Machina Mirabilis'' \supercite{hla2026machinamirabilis}: remove general relativity from the training distribution, provide only pre-relativistic observations plus consistency constraints, and evaluate whether the system can invent principles similar to Einstein's original insights.
\vspace{4pt}

If AI systems can discover simple theories in artificial worlds, or rediscover known theories when the historical answer is withheld, but still fail to produce comparable advances in the real world -- that failure would itself be important. It would suggest that the bottleneck may lie not only in AI architecture, but also in the intrinsic complexity or experimental untestability of the remaining undiscovered theories. 
Either outcome would be valuable. Success would show a path toward AI-assisted theory discovery, while failure under controlled conditions would teach us something profound about both AI and the present state of theoretical physics.
\end{tcolorbox}

\section{Formalizing the Paradigm Shift}
\label{sec:formalizing-shift}

If AI is to synthesize new theories, we must ask: what is the native language of physics discovery? Currently, generative AI relies predominantly on natural language, simply because of the sheer abundance of textual data. 
Natural language in the form of physical data is highly expressive, but it lacks the structure required for mathematical physics.

Conversely, the mathematics community is increasingly turning to Interactive Theorem Provers (ITPs) such as Lean \supercite{moura2015lean} to formalize proofs. Recent efforts suggest that Lean could also be adapted to formalize certain domains of physics \supercite{breen2025axproverdeepreasoningagentic}. However, traditional ITPs may only be sufficient for Category A and B discoveries -- deriving predictions and phenomenological laws from established frameworks. i.e., Lean is designed to prove that a conclusion follows from fixed axioms, but it is not designed to realize that the axioms themselves need to be replaced.

Therefore, bringing formalization to Category C discovery will require a fundamentally new type of formal system: a ``Reverse ITP''. 
Instead of navigating from fixed axioms to a conclusion, this system should formalize the abductive process itself as an organized search over provisional principles.
A generative AI system would propose ``temporary axioms'' -- bold physical ansätze, symmetries, or other mathematical structures -- and use symbolic tools to deduce their falsifiable consequences. It would then test these consequences against data or consistency constraints, modifying parameters or the provisional axioms themselves when contradictions arise.
In this sense, theoretical contradictions become analogous to a loss signal for refining the provisional axioms of the candidate framework.
The output would not be a proof alone, as in conventional ITPs, but a set of candidate frameworks, each with provisional axioms and derived consequences that can be automatically tested and falsified.

Human scientists communicate ideas through representational languages that trade precision for expressive range. At one end is formal mathematics: equations, definitions, and theorems can be transmitted with very high precision, but only after an idea has been cast into a sufficiently rigid form. Natural language occupies a middle ground: it can express hypotheses, analogies, intuitions, and incomplete arguments, but at the cost of increased ambiguity. At the other end are artistic modes of communication, such as images, music, and metaphor, which can convey broader forms of intuition and experience, but with even less determinate meaning. 
Physics has historically relied on all parts of this spectrum. Although its final results are often expressed mathematically, the reasoning that leads to them often involves ``controlled vagueness'': diagrams, analogies, ansätze, scaling arguments, effective descriptions, and provisional concepts whose meaning becomes sharper only through use. 

This line of thinking suggests that a Lean-like formal system may be too rigid for physics, especially in the early stages of theory formation. Physics may need a representational medium that is perhaps just a bit more flexible than formal mathematics but clearly more structured than ordinary natural language or art. The relevant goal may therefore be a semi-formal language of physics discovery: precise enough to support symbolic manipulation, consistency checks, and derivation of falsifiable consequences, yet flexible enough to represent approximate principles, incomplete analogies, emergent variables, and candidate mathematical structures before they have been fully formalized. Such a language may need to preserve something art-like: the ability to communicate rich, partially formed intuitions before they have been compressed into equations.

If AI is to contribute to paradigm-shifting physics, we should help teach it the mode of discovery that shaped modern physics from Einstein to Higgs (and since): 
not just fitting patterns, 
but constructing principle-constrained frameworks and deducing their falsifiable consequences. 
Only then will we finally know whether AI can actually discover the next Standard Model, or merely keep predicting around it.

\newpage
\printbibliography

\end{document}